\documentclass[pre,nofootinbib,amsmath,amssymb,preprint,showpacs]{revtex4}
\usepackage[T1]{fontenc} \usepackage[latin1]{inputenc}
\usepackage{amsmath} \usepackage{graphicx} \usepackage{amssymb}
\usepackage{subfigure,epsfig,color}

\begin{document}

\title{Dynamical Eigenmodes of Star and Tadpole Polymers}

\author{Rick Keesman} \affiliation{Institute for
  Theoretical Physics, Universiteit Utrecht, Leuvenlaan 4, 3584 CE
  Utrecht, The Netherlands} \author{Gerard T. Barkema} \affiliation{Institute for
  Theoretical Physics, Universiteit Utrecht, Leuvenlaan 4, 3584 CE
  Utrecht, The Netherlands} \affiliation{Instituut-Lorentz,
  Universiteit Leiden, Niels Bohrweg 2, 2333 CA Leiden, The
  Netherlands} \author{Debabrata Panja} \affiliation{Institute of
  Physics, Universiteit van Amsterdam, Valckenierstraat 65, 1018 XE
  Amsterdam, The Netherlands} \affiliation{Institute for Theoretical
  Physics, Universiteit Utrecht, Leuvenlaan 4, 3584 CE Utrecht, The
  Netherlands}

\begin{abstract}
  The dynamics of phantom bead-spring chains with the topology of a
  symmetric star with $f$ arms and tadpoles ($f=3$, a special case) is
  studied, in the overdamped limit. In the simplified case where the
  hydrodynamic radius of the central monomer is $f$ times as heavy as
  the other beads, we determine their dynamical eigenmodes exactly,
  along the lines of the Rouse modes for linear bead-spring
  chains. These eigenmodes allow full analytical calculations of
  virtually any dynamical quantity. As examples we determine the
  radius of gyration, the mean square displacement of a tagged
  monomer, and, for star polymers, the autocorrelation function of the
  vector that spans from the center of the star to a bead on one of
  the arms.
\end{abstract}


\maketitle

\section{Introduction\label{sec:intro}}

Bead-spring models play a central role in the theory and modelling of
polymer dynamics. Most of the applications of bead-spring models (and
polymer dynamics in general) is found for linear polymers, for which
the polymer consists of a linear sequence of beads connected by
harmonic springs. For a linear bead-spring polymer chain, with
position ${\bf R}_n$ of the $n$-th bead, $n=0\dots N$, the potential
energy is thus
\begin{equation}
U= \frac{k}{2} \sum_{n=1}^N \left({\bf R}_n - {\bf R}_{n-1}
\right)^2,
\label{e1}
\end{equation}
in which $k$ is the spring constant. Besides the heavily studied
linear polymer chains there are many other types of polymers whose
dynamics deserve closer inspection. In this paper we concern ourselves
with the dynamics of bead-spring chains that have the topology of a
symmetric star (with $f$ arms) and tadpoles.

Further, in the context of bead-spring models, the Rouse model for
linear phantom chains \cite{rouse} deserves a special mention. In the
Rouse model the dynamics of the beads is formulated in the overdamped
limit, with a solvent with viscous drag coefficient $\zeta$, and with
thermal forces ${\bf g}_n$ on the $n$-th bead, such that the dynamical
equations of motion become
\begin{equation}
\frac{d {\bf R}_n}{dt}= - \frac{1}{\zeta} \frac{\partial U}{\partial
  {\bf R}_n} + {\bf g}_n.
\label{rouseeq}
\end{equation}
Here, the thermal forces are delta-correlated, namely
\begin{equation}
\label{eq:thermal}
\left< {\bf g}_i(t) \cdot {\bf g}_j(t^\prime) \right> = \frac{6 k_B T}{\zeta} \delta_{ij}\delta(t-t^\prime),
\end{equation}
with Boltzmann constant $k_B$ and temperature $T$.

The reason why, to date, the Rouse model deserves a special mention,
lies in the fact that it allows
full analytical calculations of virtually any dynamical
quantity. Rather than the equations of motion for the individual
beads, one considers the so-called Rouse modes, whose amplitudes at
time $t$ are given by
\begin{equation}
\label{eq:chainmode}
{\bf X}_p(t) = \frac{1}{N+1}\sum_{n=0}^N\cos\left[\frac{\pi(n+1/2)p}{N+1}\right]\, {\bf R}_n(t),
\end{equation}
with $p=0 \dots N$. The dynamics of these modes are uncorrelated, and
one can derive the following relation for $p,q\neq0$ exactly \cite{doi,de} from the
equation of motion (\ref{rouseeq}):
\begin{eqnarray}
X_{pq}(t)\equiv\langle {\bf X}_p(t)\cdot {\bf X}_q(0)\rangle=A_1\frac{N}{p^2}\,\exp\left[-A_2\,\frac{p^2}{N^2}\,t\right]\,\delta_{pq},
\label{eq:chainmacf}
\end{eqnarray}
where $A_1=3 k_B T/(2 \pi^2 k)$ and $A_2=\pi^2 k/\zeta$. Here, and all
throughout this paper, the angular brackets represent an average over
the equilibrium ensemble of polymer chains. Equation
(\ref{eq:chainmacf}) is further supplemented by $X_{0p}(t)=0$ for
$p\neq0$, and $X_{00}(t)\equiv\langle [{\bf X}_0(t) - {\bf
  X}_0(0)]^2\rangle=6 k_B T t/(\zeta N)$, where ${\bf X}_0(t)$ is
the location of the center-of-mass of the polymer at time $t$.  Using
these mode amplitude correlation functions, the quantities of interest
for a phantom chain can be analytically tracked by reconstructing them
from the modes \cite{doi,de}.

Using the definition of Rouse modes it has been shown in recent works
that the dynamics of a tagged bead in a linear bead-spring model is
described by the Generalized Langevin Equation (GLE)
\cite{jstat1,jstat2}, and that the dynamics of polymers with steric
repulsion (also known as self-avoiding polymer chains)
\cite{saw-rouse}, as well as of those in melts as described by the
repton model \cite{reptonmodes} can be well-approximated. Here, we
continue this line of research, but now we are interested in exact
solutions of the dynamical properties of polymers with the
topology of stars and tadpoles.

The structure of this paper is as follows. In Sec. \ref{sec:star} we
present the dynamical eigenmodes of symmetric phantom star polymers,
and use the mode amplitudes to provide analytical expressions to the
radius of gyration, mean square displacement of a tagged monomer, and
the autocorrelation function of the vector that spans from the center of the
star to a bead on one of the arms. In Sec. \ref{sec:tadpole} we
repeat the exercise for chains with tadpole topology. We end the paper
with a short discussion in Sec. \ref{disc}.

\section{Dynamical eigenmodes of symmetric star
  polymers \label{sec:star}}

\begin{figure}[h]
  \begin{center}
    \epsfig{file=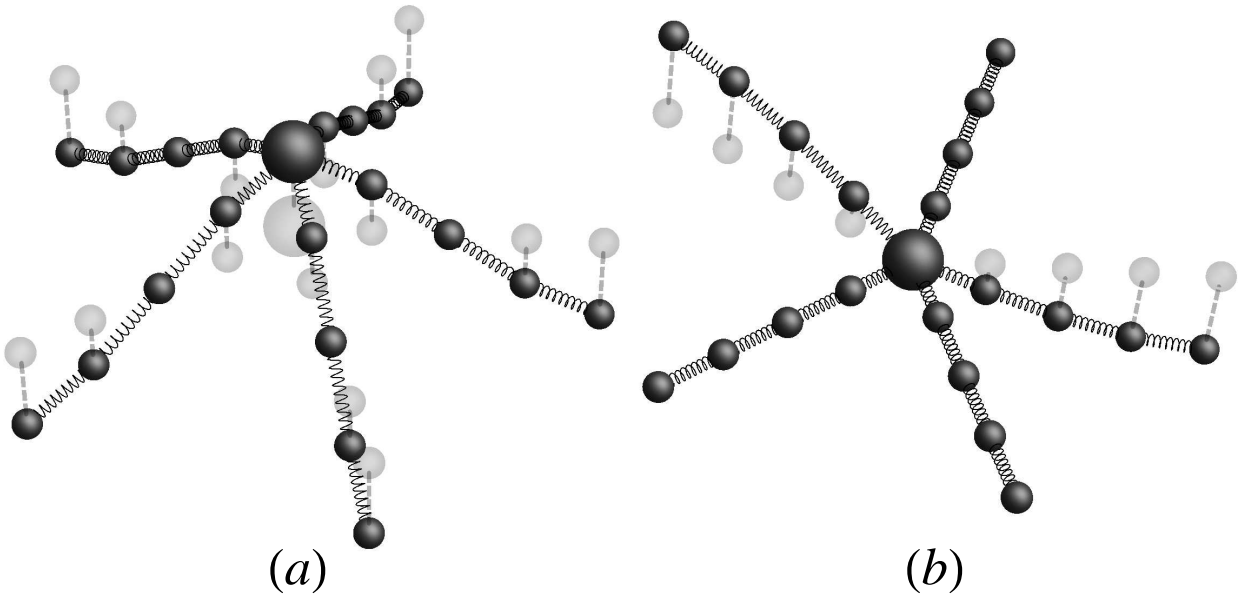,width=1.0\linewidth,clip=}
  \end{center}
  \caption{Two graphical representations of a symmetrical star polymer
    to visualize the Rouse modes. Depicted is a symmetric starpolymer
    with five arms consisting of four beads each and a central monomer
    which has a friction coefficient five times that of beads in the
    arms.  The transparent configurations are polymers in the origin
    stretched in the $xy$-plane for visual convenience with no Rouse
    mode excited in the $z$-direction.  The opague configurations are
    like the transparent ones but with a pure ${\bf X}_1$ mode (a) and
    ${\bf Y}_{1}^{(i,j)}$ modes (b) in the z-direction.}
  \label{fig:star}
\end{figure}

A major difficulty for dynamics of polymers with a more complex
topology like the symmetric star polymer is that in most cases an
elegant analytical expression for the set of dynamical eigenmodes
cannot be found \cite{starthesis}. Here we show that for a symmetric
star polymer with a special central bead the dynamical eigenmodes can
indeed be written down exactly, and subsequently the dynamical
behavior of many interesting physical quantities can be determined
precisely. For simplicity, in commensuration with the Rouse model,
henceforth we term these dynamical eigenmodes as Rouse
modes. Specifically, we consider a star polymer with $f$ identical
arms, each consisting of $N$ identical beads, connected to a central
bead whose hydrodynamic radius is $f$ times as big as the other ones,
i.e., its viscous drag is $f$ times as large. A graphical
representation of such a star polymer with $f=5$ and $N=4$ can be
found in Fig. \ref{fig:star}. It is worthwhile to note in this
context that a key method to synthesize a star polymer chain is to
attach the arms, which are linear chains, to a multivalent central
core with sticky ends (see, e.g., Ref. \cite{gao} and the references
cited therein). Although from the synthesis process it is realistic
that the core has a significantly higher hydrodynamic radius, the
choice in our simplified model to make hydrodynamic radius of the
central bead exactly $f$ times as the other beads is motivated by our
strive to determine the Rouse modes exactly. For modestly-sized
chains, where the hydrodynamic radius of the core is not $f$ times as
big as the arm beads, the dynamical matrix (the homogenous part of the
dynamical differential equation) has been diagonalized numerically,
yielding the numerical identification of the Rouse modes
\cite{starthesis}.

We label the position of the central bead as ${\bf R}_0$ and let
${\bf R}_{a,n}$ be the position of the $n$-th bead, $n=1\dots N$, in the
$a$-th arm, $a=1\dots f$. We consider two types of Rouse modes given
by
\begin{subequations}
\label{eq:starmode}
\begin{eqnarray}
\label{eq:starxmode}
{\bf X}_{p}(t) &=& \frac{1}{N+1}\left\{\cos\left[\frac{\pi p/2}{N+1}\right]{\bf R}_{0}(t) + \frac{1}{f}\sum_{a,n=1}^{f,N} \cos\left[\frac{\pi(n+1/2)p}{N+1}\right] {\bf R}_{a,n}(t) \right\}, \\
\label{eq:starymode}
{\bf Y}_{q}^{(i,j)}(t)&=&\frac{1}{2N+1} \sum_{n=1}^{N} \cos\left[\frac{\pi(N-n+1/2)(q-1/2)}{N+1/2}\right] \left( {\bf R}_{i,n}(t) - {\bf R}_{j,n}(t) \right),
\end{eqnarray}
\end{subequations}
with $p=0 \dots N$ and $q=1 \dots N$. A visualization of the two kinds
of modes can also be found in Fig. \ref{fig:star}. The first set of
modes ${\bf X}_{p}(t)$ are like Rouse modes for a stringpolymer
through all the arms. The second set of modes ${\bf Y}_{p}^{(i,j)}(t)$
can also be thought of as Rouse modes as in Eq.  (\ref{eq:chainmode})
with $p$ odd valued and through a linear chain of length $2 N + 1$
made up by arms $i$ and $j$ and the central bead. There are
$Nf(f-1)/2$ modes of the type ${\bf Y}_{p}^{(i,j)}(t)$ with $i<j$, but
the total set of these modes contain only $(f-1)N$ independent degrees
of freedom, for instance because ${\bf Y}_{p}^{(i,k)}(t)$=${\bf
  Y}_{p}^{(i,j)}(t)$+${\bf Y}_{p}^{(j,k)}(t)$; we could have
constructed, for every $p$, an orthogonal set of $f-1$ modes out of
the full set of modes ${\bf Y}_{p}^{(i,j)}(t)$, but we choose not to
do that for the sake of mathematical elegance.  Combined with $N+1$
modes ${\bf X}_{p}(t)$, the total set contains $f N +1$ three-dimensional modes needed
to describe the system that has $f N +1$ beads so that the number of
degrees of freedom is the same.

Similarly to the single chain, the dynamics of these modes for a
star polymer with long arms are captured by
\begin{subequations}
\label{eq:starmacf}
\begin{eqnarray}
\label{eq:starxmacf}
X_{pq}(t)&\equiv&\langle {\bf X}_p(t)\cdot {\bf X}_q(0)\rangle=\frac{3 k_b T}{2 \pi^2 k}\frac{N}{f p^2}\,\exp\left[-\frac{\pi^2 k}{\zeta}\,\frac{p^2}{N^2}\,t\right]\,\delta_{pq}, \\
\label{eq:starymacf}
Y_{pq}^{(i,j)(k,l)}(t) &\equiv&\langle {\bf Y}_{p}^{(i,j)}(t) \cdot {\bf Y}_{q}^{(k,l)}(0) \rangle \nonumber \\
&=& \frac{3 k_b T}{8 \pi^2 k} \frac{N}{(p-1/2)^2} \, \exp \left[ -\frac{\pi^2 k}{\zeta} \frac{(p-1/2)^2}{N^2} t \right] \delta_{p q} \delta_{(i,j)(k,l)}  ,
\end{eqnarray}
\end{subequations}
where $\delta_{(i,j)(k,l)}=\delta_{i k}-\delta_{j k}-\delta_{i
  l}+\delta_{j l}$. This is supplemented by $X_{00}(t)\equiv\langle [
{\bf X}_{0}(t) - {\bf X}_{0}(0) ]^2 \rangle = 6 k_B Tt/(\zeta f (N+1))$
and all other correlations between modes strictly zero. The
derivations of the Rouse mode amplitudes can be found in Appendix A.

\subsection{Radius of gyration}
The squared radius of gyration is defined as the weighted sum over all
differences between the position of a monomer and the center of mass
squared. Below we work out the radius of gyration for the case of the
star polymer where the central monomer is $f$ times heavier than the
other beads, although the radius of gyration can be calculated
following the same line for other cases as well. In the former case
the location of the center-of-mass ${\bf R}_{cm}(t)\equiv {\bf
  X}_{0}(t)$, and the squared radius of gyration is defined as
\begin{equation}
    \label{eq:gyration1}
    R_{g}^{2}  = \frac{1}{f(N+1)} \left[ f \langle [ {\bf R}_{0}(t)- {\bf R}_{cm}(t) ]^{2} \rangle + \sum_{i,n=1}^{f,N} \langle [ {\bf R}_{i,n}(t)- {\bf R}_{cm}(t) ]^{2} \rangle \right],
\end{equation}
which can be calculated by plugging in Eq. (\ref{eq:starinv}). This reduces to
\begin{align}
    \label{eq:gyration3}
    R_{g}^{2} = 2\sum_{p=1}^{N} X_{p p}(0) + \frac{4(2 N +1)}{f^{3}(N+1)} \sum_{p,i,j,k=1}^{N,f,f,f} Y_{p p}^{(i,j)(i,k)}(0).
\end{align}
Plugging in Eq. (\ref{eq:starmacf}) explicitly and taking the long polymer limit yields
\begin{align}
    \label{eq:gyration4}
    R_{g}^{2} = \frac{3 k_B T}{k \pi^2} \frac{N}{f} \sum_{p=1}^{\infty} \frac{1}{p^{2}} +
    \frac{3 k_B T}{k \pi^2} \frac{N}{f^3} \sum_{i,j,k=1}^{f,f,f} \delta_{(i,j)(i,k)} \sum_{p=1}^{\infty} \frac{1}{(p-1/2)^{2}}.
\end{align}
The sums can be evaluated by using Eq. (\ref{eq:sums}).

And so the radius of gyration squared for a long symmetric star polymer becomes
\begin{align}
    \label{eq:gyration6}
    R_{g}^{2} &= \frac{k_B T N}{k} \frac{3 f - 2}{2f}.
\end{align}
This result is consistent with that of a linear chain ($f=1$).

\subsection{Mean square displacement of a monomer}

\begin{figure*}
    \centering
    \epsfig{file=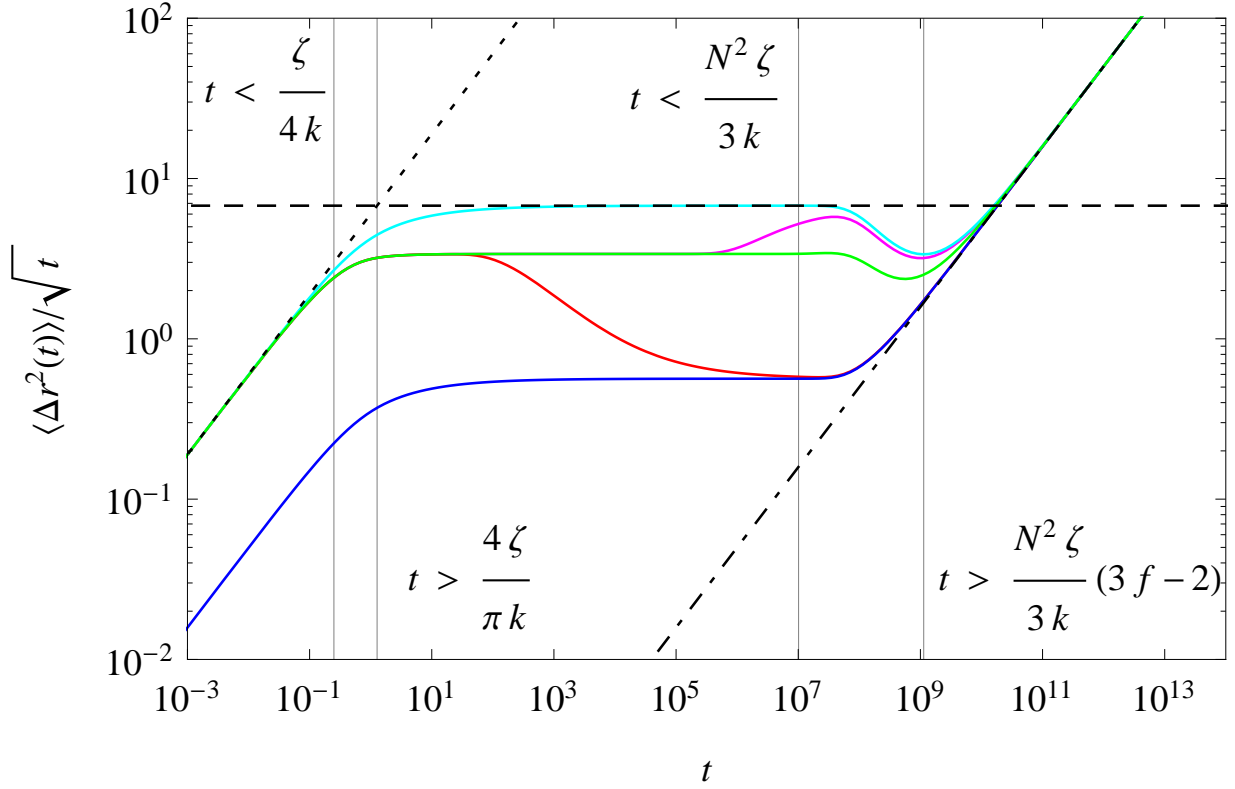,width=1.0\linewidth,clip=}
    \caption{The scaled mean square displacement $\langle \Delta {\bf
        R}^{2}(t) \rangle/\sqrt{t}$ as a function of time. The
      mean square displacement of several beads in a symmetric star
      polymer as given by Eqs.  (\ref{eq:starmsd2},\ref{eq:starmsd6})
      were exactly evaluated with $A=12$ and $N=10^4$ and other
      parameters put to $1$. The solid blue line corresponds to the
      mean square displacement of the central bead whereas the red,
      green, magenta, and cyan solid lines correspond to $n=10^{-3} N,
      N/2, 9N/10$, and $N$ in Eq. (\ref{eq:starmsd6})
      respectively. A bead positioned somewhere along the arm will at
      first behave as if it were in the middle of the arm. After time
      grows either the end or the center will become an influence at
      first and the bead will start mimicking a bead at one of those
      places. The short time scale $t < \zeta/(4 k)$ and the very long
      time scale $t \gg N^2 \zeta(3f-2)/(3 k)$ for which $\langle
      \Delta {\bf R}^{2}(t) \rangle \sim t$ corresponding to the
      dotted and dot-dashed lines are separated by a time during which
      $\langle \Delta {\bf R}^{2}(t) \rangle \sim \sqrt{t}$
      corresponding to the dashed line as in agreement with Eqs.
      (\ref{eq:starmsd4},\ref{eq:starmsd8}-\ref{eq:starmsd9}).}
    \label{fig:starmsd}
\end{figure*}

We now consider the mean square displacement of the central bead and a
bead in an arm for a star polymer with long arms. We start by
defining the displacement vector for the central bead and writing it in
terms of modes using Eq. (\ref{eq:starinva}):
\begin{align}
    \label{eq:starmsd1}
    \Delta {\bf R}_{0}(t)  \equiv & {\bf R}_{0}(t) - {\bf R}_{0}(0) \nonumber\\
      = & {\bf X}_{0}(t) - {\bf X}_{0}(0) + 2 \sum_{p=1}^{N} \cos\left[\frac{\pi p/2}{N+1}\right] \left\{{\bf X}_{p}(t) - {\bf X}_{p}(0) \right\}  .
\end{align}
The mean square displacement is then given by
\begin{align}
    \label{eq:starmsd2}
    \langle \Delta {\bf R}_{0}^{2}(t) \rangle  = X_{0 0}(t) + 8 \sum_{p=1}^{N} \cos^2\left[\frac{\pi p/2}{N+1}\right] \{X_{p p}(0) - X_{p p}(t)\},
\end{align}
where Eq. (\ref{eq:starxmacf}) can be plugged in, and the
orthogonality of the modes was already used for simplification. Before
doing so, let us look at very short time scales $t <\zeta/(4 k)$. The
exponent in $X_{pq}(t)$ can then be expanded and the sum exactly
evaluated using Eq. (\ref{eq:minsum1}), resulting in $\langle
\Delta {\bf R}_{0}^{2}(t) \rangle = 6 k_B Tt/(\zeta f)$ dominating
over the mean square displacement of the whole polymer which at short
time scales is negligible. At very long time scales $X_{pq}(t)$ goes
to zero and the term with the summation in Eq.
(\ref{eq:starmsd2}) can again be exactly evaluated using Eq.
(\ref{eq:maxsum1}). So for $t<\zeta N^2/(3 k)$ the summation has a
larger contribution than the mean square displacement of the whole
polymer. For intermediate times the summation dominates and
can be rewritten as an integral for very long polymers:
\begin{align}
    \label{eq:starmsd3}
    \langle \Delta {\bf R}_{0}^{2}(t) \rangle  =
     \frac{12 k_B T}{\pi^2 k f} \int_{0}^{\infty} \frac {dx}{x^{2}} \left\{ 1 - \exp \left[ - \frac{k \pi^2 t}{\zeta} x^2 \right] \right\} = \frac{12 k_B T}{f}\sqrt{\frac{t}{\pi k \zeta}}.
\end{align}
The mean square displacement as approximated for intermediate times
is greater than that of the approximation for the short time
scales for $t>4 \zeta/(\pi k)$, so that is when the intermediate time
regime begins. For the central bead in a star polymer the mean square
displacement then becomes
\begin{equation}\label{eq:starmsd4}
\langle \Delta {\bf R}_{0}^{2}(t) \rangle =
\left\{
	\begin{array}{ll}
        \displaystyle{\frac{6 k_B T}{\zeta f} t,
          \quad\quad\quad\quad\quad  \mbox{for}\, t<\frac{\zeta}{4 k}}\\
        \displaystyle{\frac{12 k_B T}{f}\sqrt{\frac{t}{\pi k \zeta}},
          \quad\quad  \mbox{for}\, \frac{4 \zeta}{\pi k} < t < \frac{N^2 \zeta}{3 k}}\\
        \displaystyle{\frac{6 k_B T}{\zeta f (N+1)} t, \quad\quad\quad\quad\quad  \mbox{for}\,  t > \frac{N^2 \zeta}{3 k}}
	\end{array}
\right..
\end{equation}
Figure \ref{fig:starmsd} shows the exact evaluation of Eq.
(\ref{eq:starmsd2}) for some star polymer together with approximations
made for the short, intermediate and long time scales found in Eq.
(\ref{eq:starmsd4}). The central bead first behaves like a single bead
with friction coefficient $\zeta f$. After that the movement is
restricted by local connections to surrounding beads in the
polymer. For very long times the position of the bead within the polymer
is negligible and the mean square displacement behaves as that of a
single bead with friction coefficient $\zeta f (N+1)$.

A very similar approach can be used for the mean square displacement
of a bead in an arm of the star polymer, defined as
\begin{align}
    \Delta {\bf R}_{n}(t) \equiv & {\bf R}_{i,n}(t) - {\bf R}_{i,n}(0) \nonumber\\ =
    &{\bf X}_{0}(t) - {\bf X}_{0}(0) +
    2\sum_{p=1}^{N} \cos\left[\frac{\pi (n+1/2) p}{N+1}\right] \{{\bf X}_{p}(t) - {\bf X}_{p}(0)\} \nonumber\\
    &+\frac{4}{f} \sum_{j,p=1}^{f,N} \cos \left[ \frac{\pi (N-n+1/2) (p-1/2)}{N+1/2}\right]  \{{\bf Y}^{(i,j)}_{p}(t) - {\bf Y}^{(i,j)}_{p}(0) \}.     \label{eq:starmsd5}
  \end{align}
Using orthogonality of mode ${\bf Y}^{(i,j)}_{p}$ with ${\bf X}_{q}$ and
with ${\bf Y}^{(k,l)}_{q}$ with $q \neq p$, and evaluating the double
sum over the arms, the mean square displacement becomes
\begin{align}
    \langle \Delta {\bf R}_{n}^{2}(t) \rangle  = & X_{00}(t) +
     8 \sum_{p=1}^{N} \cos^{2}\left[\frac{\pi (n+1/2) p}{N+1}\right]  \left\{ X_{p p}(0) - X_{p p}(t)\right\} \nonumber \\
    & +\frac{16(f-1)}{f} \sum_{p=1}^{N}  \cos^{2} \left[ \frac{\pi (N-n+1/2) (p-1/2) }{N+1/2}\right]  \nonumber \\
    &\times\left\{ Y_{p p}^{(1,2)(1,2)}(0) - Y_{p
        p}^{(1,2)(1,2)}(t)\right\} .
    \label{eq:starmsd6}
\end{align}
For small values of $t$ the exponents can again be expanded and the
sum exactly evaluated using Eqs. (\ref{eq:minsum1}-\ref{eq:minsum2}) so
that in the long polymer limit $ \langle \Delta {\bf R}_{n}^{2}(t)
\rangle = 6 k_B Tt/\zeta$. The summation reaches its maximum, using
Eqs. (\ref{eq:maxsum1}-\ref{eq:maxsum2}), on long time scales so
that for a bead at the end of an arm the mean square displacement term
of the whole polymer will start to dominate for $t>N^2 \zeta (3f-2)/(3
k)$. The closer a bead is to the central bead the faster its
mean square displacement will behave like that of the whole
polymer. We can again approximate the summation by an integral for
intermediate times but depending on the position of the bead in an arm it
will behave differently.
\begin{align}
    \langle \Delta {\bf R}_{n}^{2}(t) \rangle  = &
     \frac{12 k_B T}{\pi^2 k f} \int_{0}^{\infty} \frac {dx}{x^{2}} \left\{ 1 - \exp \left[ - \frac{k \pi^2 t}{\zeta} x^2 \right] \right\}  \nonumber\\
     &\times\left\{ \cos \left[ \frac{\pi (n+1/2) p}{N+1} \right] + (f-1)\cos \left[ \frac{\pi (N-n+1/2) (p-1/2)}{N+1/2} \right] \right\}
    \label{eq:starmsd7}
\end{align}
The exponent in the integral will suppress the contribution for higher
$p$-values. For a bead at the end of an arm, $n$ equals $N$, the cosines can
be taken to be $1$ and the integral as the same as the integral for
the central bead but with an extra factor $f$. For a bead in the
middle of an arm, $n$ equals $N/2$, in the long polymer limit the first cosine
will be $1$ for even values of $p$ and $0$ for odd values, and the
second cosine will be $1/2$ for all values of $p$. This will give the
same integral as for the bead at the end of an arm but smaller by a factor
$1/2$. We thus find for the mean square displacement of a bead
at the end of an arm
\begin{equation}\label{eq:starmsd8}
\langle \Delta {\bf R}_{N}^{2}(t) \rangle =
\left\{
	\begin{array}{ll}
          \displaystyle{\frac{6 k_B T}{\zeta} t, \quad \quad \quad
            \quad \quad \mbox{for}\, t<\frac{\zeta}{4 k}}\\
          \displaystyle{ 12 k_B T\sqrt{\frac{t}{\pi k \zeta}}, \quad \quad \mbox{for}\, \frac{4 \zeta}{\pi k} < t < \frac{N^2 \zeta}{3 k}}\\
          \displaystyle{\frac{6 k_B T}{\zeta f N} t, \quad \quad \quad \quad \quad\mbox{for}\, t > \frac{N^2 \zeta}{3 k}(3f-2)}
	\end{array}
\right..
\end{equation}
For the bead located exactly in the middle of an arm the mean square
displacement becomes
\begin{equation}\label{eq:starmsd9}
\langle \Delta {\bf R}_{N/2}^{2}(t) \rangle =
\left\{
	\begin{array}{ll}
          \displaystyle{\frac{6 k_B T}{\zeta} t,\quad \quad \quad
            \quad \quad \mbox{for}\,  t<\frac{\zeta}{4 k}}\\
          \displaystyle{ 6 k_B T\sqrt{\frac{t}{\pi k \zeta}}, \quad \quad \,\,\,\,\mbox{for}\, \frac{4 \zeta}{\pi k} < t < \frac{N^2 \zeta}{3 k}}\\
          \displaystyle{\frac{6 k_B T}{\zeta f N} t, \quad \quad \quad \quad \quad\mbox{for}\, t > \frac{N^2 \zeta}{3 k}(3f-2)}
	\end{array}
\right..
\end{equation}
Figure \ref{fig:starmsd} shows the exact evaluation of Eq.
(\ref{eq:starmsd6}) and demonstrates the validity of Eqs.
(\ref{eq:starmsd8}-\ref{eq:starmsd9}). The behavior of a bead
somewhere along the arm can also be extracted from these results. At
first it will behave like the bead in the middle of an arm since
locally they are the same. As time progresses it will either start
feeling the end or the center of the polymer at first and mimic the
behavior of the bead at the end or center respectively after which the
mean square displacement will behave like that of the whole polymer.

\subsection{Correlation function of a vector connecting a bead to the
  central bead}

Consider the spatial vector connecting a bead in some arm to the
central bead
\begin{align}
    \label{eq:starsv1}
    {\bf r}_{i,n}(t)  \equiv {\bf R}_{i,n}(t) - {\bf R}_{0}(t) .
\end{align}
\begin{figure*}
    \centering
    \epsfig{file=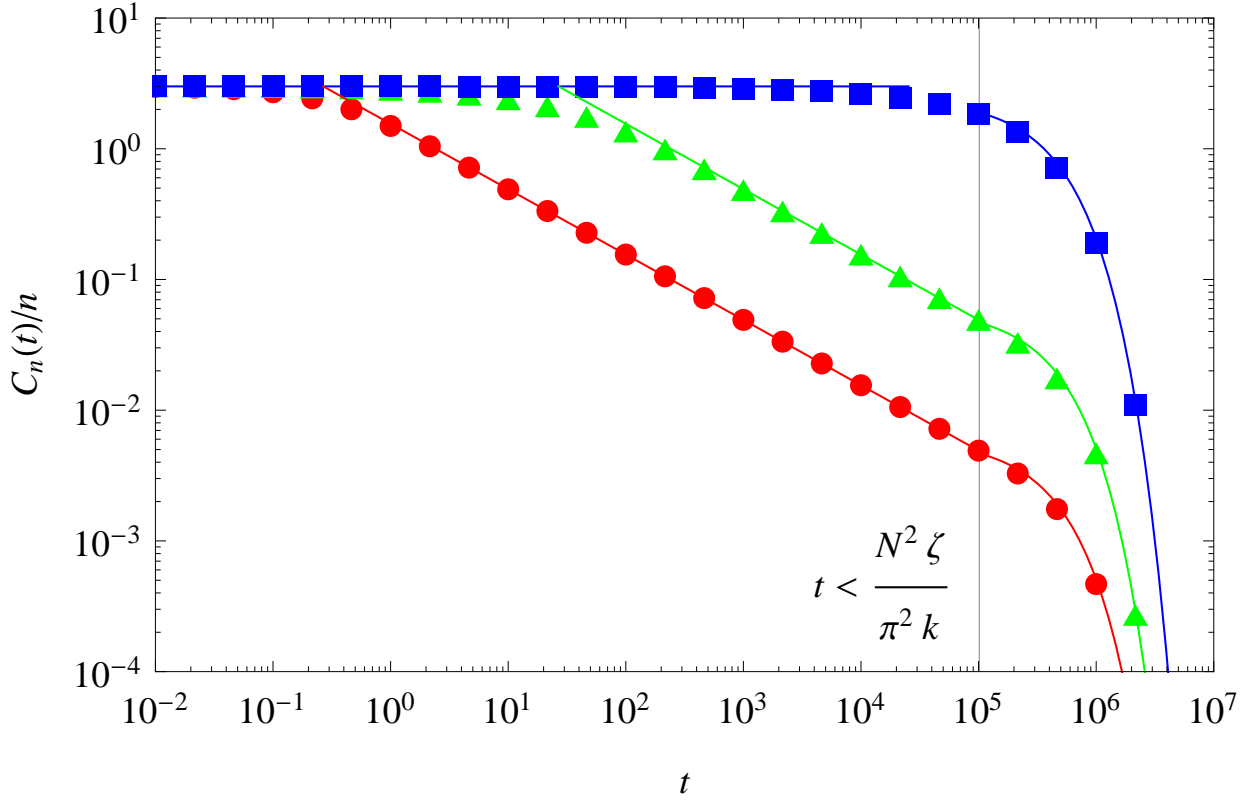,width=1.0\linewidth,clip=}
    \caption{The scaled correlation function $C_{n}(t)/n$ as defined in
      Eqs. (\ref{eq:starsv1}-\ref{eq:starsv2}) in a double-logarithmic
      plot as a function of time. The correlation functions for
      several monomers in a symmetric star polymer as given by
      Eq. (\ref{eq:starsv3}) were exactly evaluated with $f=12$ arms
      of length $N=10^3$ and other parameters put to $1$. The red disks,
      green triangles, and blue squares correspond to $n=1,10$, and $N$ for $C_{n}(t)$
      respectively. Approximations for the correlation function in the
      short time scale $t < \zeta(n+1)^2/(4 \pi^2 k)$, intermediate
      time scale $\zeta n^2 (f-1)^2/(k \pi f^2) < t < N^2 \zeta/(\pi^2
      k)$, and the very long time scale $t > N^2 \zeta/(\pi^2 k)$ as
      in Eq. (\ref{eq:starsv5}) correspond solid lines. The approximation for intermediate
      times is only very accurate for small $n$, but for larger
      $n$ this time domain becomes smaller or even non-existent as is
      the case for $n$ equals $N$ as can be seen in the figure.}
    \label{fig:starsv}
\end{figure*}
The correlation function $C_{n}(t) \equiv\langle {\bf r}_{i,n}(t)\cdot {\bf
  r}_{i,n}(0)\rangle$ is then given by
\begin{align}
    C_{n}(t) =& 16 \sum_{p=1}^{N} \sin^2 \left[ \frac{\pi (n+1)p/2}{N+1} \right] \sin^2 \left[ \frac{\pi n p/2}{N+1} \right] X_{p p}(t) \nonumber \\
    & +\frac{16}{f^2} \sum_{j,k,p=1}^{f,f,N} \cos^2 \left[ \frac{\pi (N-n+1/2)(p-1/2)}{N+1/2} \right] Y_{p p}^{(i,k)(i,j)}(t) ,
    \label{eq:starsv2}
\end{align}
where the cosines were reduced to sines for notational
convenience. Having filled in the mode dynamics functions explicitly
yields
\begin{align}
    \label{eq:starsv3}
    C_{n}(t) =
    & \frac{24 k_B T}{\zeta f (N+1)} \sum_{p=1}^{N} \sin^2 \left[ \frac{\pi (n+1)p/2}{N+1} \right] \sin^2 \left[ \frac{\pi n p/2}{N+1} \right] \frac{1}{\alpha_{{\bf X}_{p}}} \exp\left[ - \alpha_{{\bf X}_{p}} t \right]\nonumber \\
    & +\frac{12(f-1) k_B T}{\zeta f (2N+1)} \sum_{p=1}^{N} \cos^2 \left[ \frac{\pi (N-n+1/2)(p-1/2)}{N+1/2} \right] \frac{1}{\alpha_{{\bf Y}_{p}}} \exp\left[ - \alpha_{{\bf Y}_{p}} t \right] ,
\end{align}
where $\alpha_{{\bf X}_{p}}$ and $\alpha_{{\bf Y}_{p}}$ are defined in
Eqs. (\ref{eq:starxeom}-\ref{eq:staryeom}).  At very large and very
small times the exact correlation functions for the modes in Eq.
(\ref{eq:starmacfexact}) can be used. At very small times the
exponential can be omitted and the resulting equations solved using
Eqs.  (\ref{eq:maxsum2},\ref{eq:maxsum4}). Adding the two terms then
results in $C_{n}(t) = 3 n k_B T/k$. To determine the region
for which this approximation is valid we notice that the exponent is
roughly $1$ for $p^2 < \zeta N^2/(\pi^2 k t)$. The term in the
summation will go to $0$ for $p=2(N+1)/(n+1)$ and so the
largest contribution to the summation is for all the terms before this
happens. Solving this for $t$ where the large polymer limit is taken
gives $t<\zeta (n+1)^2/(4 \pi^2 k)$ for which the approximation
is valid. For very large time values $t>(N^2 \zeta)(\pi^2 k)$
only the lowest mode will give contribution. Since $f=1,2$ are two
cases of linear chains we focus on $f\geq3$ for which the second term
in Eq. (\ref{eq:starsv3}) will dominate. Taking the long polymer
limit the sine can be expanded and the cosine reduced for notational
convenience. For very large times the correlation function can thus be
approximated by $C_n(t) = 12 k_B T N/(\pi^2 k)\left( 1-
  \cos\left[\pi n/N \right] \right)\exp\left[ -\pi^2
    kt/(4 \zeta N^2)\right]$. For intermediate time values the second
term can be approximated by a Gaussian integral in the long polymer
limit by expanding the cosine for small $n$.
\begin{align}
    \label{eq:starsv4}
    C_{n}(t) =
    & \frac{6(f-1) n^2 k_B T}{k f} \int_{0}^{\infty} dx \exp\left[ -\frac{\pi^2 k t}{\zeta} x^2 \right] = \frac{3(f-1)n^2 k_B T}{k f \sqrt{\pi}} \left( \frac{k t}{\zeta} \right)^{-1/2}.
\end{align}
For $t>\zeta n^2 (f-1)^2/(\pi k f^2)$ the intermediate time approximation
will be smaller and thus more accurate than the approximation for
small times. Putting this together gives for $f \geq 3$
\begin{equation}
    \label{eq:starsv5}
    C_{n}(t) =
    \left\{
    	\begin{array}{ll}
            \displaystyle{\frac{3 n k_B T}{k},
              \quad\quad\quad\quad \quad\quad\quad\quad\quad\quad \quad\quad\quad\quad\quad\,\,\mbox{for}\, t<\frac{\zeta(n+1)^2}{4 \pi^2 k}}\\
            \displaystyle{\frac{3(f-1)n^2 k_B T}{k f \sqrt{\pi}} \left( \frac{k t}{\zeta} \right)^{-1/2},\, \quad \quad\quad\quad\quad\quad\quad\mbox{for}\, \frac{\zeta n^2 (f-1)^2}{k \pi f^2} < t < \frac{N^2 \zeta}{\pi^2 k}}\\
            \displaystyle{\frac{12 k_B T N}{\pi^2 k}\left( 1- \cos\left[ \frac{\pi n}{N} \right] \right)\exp\left[ -\frac{\pi^2 k}{4 \zeta N^2} t\right],\,\,\,\mbox{for}\, t > \frac{N^2 \zeta}{\pi^2 k}}
    	\end{array}
    \right..
\end{equation}
A graphical representation of the correlation function for a spatial
vector between some monomer and the central monomer in a symmetric
starpolymer with $f=12$ and $N=10^5$ is shown in Fig.
\ref{fig:starsv}.

\section{Dynamical eigenmodes of tadpole polymers
  \label{sec:tadpole}}

\begin{figure}[!ht]
  \begin{center}
    \epsfig{file=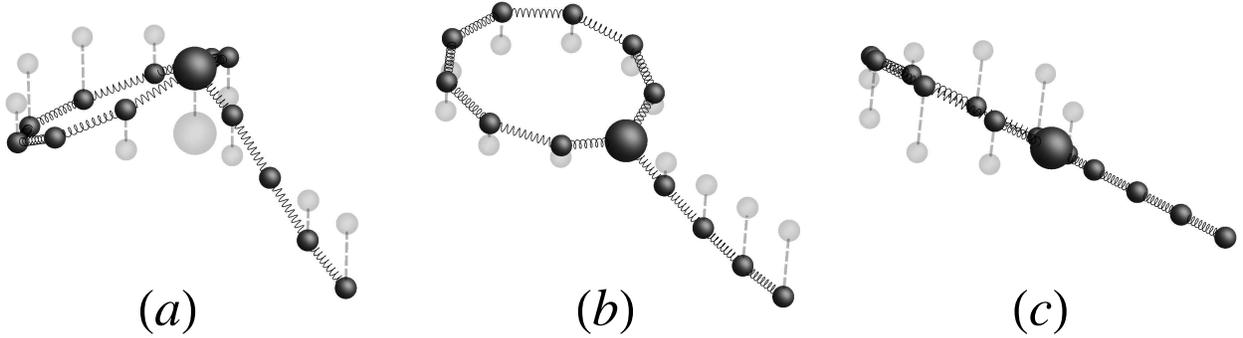,width=1.0\linewidth,clip=}
  \end{center}
  \caption{Three graphical representations of a tadpole polymer to
    visualize the Rouse modes. Depicted is a tadpole polymer, which
    can be seen as a symmetric star polymer with three arms consisting
    of four beads each where two arms are connected at the ends, with a
    central monomer which has a friction coefficient three times that of
    beads in the arms.  The transparent configurations are polymers in
    the origin stretched in the $xy$-plane for visual convenience with
    no Rouse mode excited in the $z$-direction.  The opague
    configurations are like the transparent ones but with a pure ${\bf
      X}_1$ mode (a), ${\bf Y}_{1}$ mode (b), and ${\bf Z}_1$ mode (c)
    in the z-direction.}
  \label{fig:tad}
\end{figure}

A tadpole polymer can be seen as a star polymer where two of the three
arms have the ends connected. Here we extend our calculations of the
previous section and write down the exact solution for the eigenmodes
for the specific case where the tadpole is built from a symmetric star
polymer with arms of length $N$ and a central bead with hydrodynamic
radius three times as large as that of all the other beads. The Rouse
modes are then a variation of the modes for a ring polymer combined with
those for a star polymer. The first set of $N+1$ modes are like the ${\bf X}$
modes for the star polymer where all the arms behave the same. The
second set of $N$ modes are very similar to the ${\bf Y}$ modes of the
star polymer where the ring takes on the role of an arm and the tail
of the tadpole the role of another arm. The third set ${\bf Z}$ of $N$
modes are Rouse modes for a ring polymer but specifically such that it
is antisymmetric around the central bead. In Fig. \ref{fig:tad}
these three different modes are depicted for $p=1$.
We label the tadpole polymer like a three-armed symmetric star polymer
where arm one and two have the ends connected. The Rouse modes are
then given by
\begin{subequations}
\label{eq:tadmode}
\begin{eqnarray}
\label{eq:tadxmode}
{\bf X}_{p}(t) &=& \frac{1}{N+1}\left\{\cos\left[\frac{\pi p/2}{N+1}\right]{\bf R}_{0}(t) + \frac{1}{3}\sum_{a,n=1}^{3,N} \cos\left[\frac{\pi(n+1/2)p}{N+1}\right] {\bf R}_{a,n}(t) \right\}, \\
\label{eq:tadymode}
{\bf Y}_{q}(t)&=&\frac{1}{4N+2}\!\sum_{n=1}^{N}\! \cos\!\left[\frac{\pi(N-n+1/2)(q-1/2)}{N+1/2}\right]\! \left[{\bf R}_{1,n}(t) +{\bf R}_{2,n}(t) - 2{\bf R}_{3,n}(t) \right]\!, \\
\label{eq:tadzmode}
{\bf Z}_{q}(t)&=&\frac{1}{2N+1} \sum_{n=1}^{N} \sin\left[\frac{\pi(N-n+1/2)q}{N+1/2}\right] \left( {\bf R}_{1,n}(t) - {\bf R}_{2,n}(t) \right),
\end{eqnarray}
\end{subequations}
with $p=0 \dots N$ and $q=1 \dots N$. The validity of these Rouse
modes as the dynamical eigenmodes can be proved like for the star
polymer in Appendix A. Let us define
\begin{subequations}
\begin{eqnarray}
{\bf \widetilde{X}}_{n}(t) &=& {\bf X}_{0}(t) + 2 \sum_{p=1}^{N} \cos\left[\frac{\pi (n+1/2)p}{N+1}\right] {\bf X}_{p}(t) , \\
{\bf \widetilde{Y}}_{n}(t) &=& \frac{4}{3} \sum_{p=1}^{N} \cos\left[\frac{\pi(N-n+1/2)(p-1/2)}{N+1/2}\right] {\bf Y}_{p}(t) , \\
{\bf \widetilde{Z}}_{n}(t) &=& 2 \sum_{p=1}^{N} \sin\left[\frac{\pi(N-n+1/2)p}{N+1/2}\right] {\bf Z}_{p}(t) ,
\end{eqnarray}
\end{subequations}
so that the bead locations are given by
\begin{subequations}
\label{eq:tadinv}
\begin{eqnarray}
\label{eq:tadinv0}
{\bf R}_{0}(t) &=& {\bf \widetilde{X}}_{0}(t) \\
\label{eq:tadinvx}
{\bf R}_{1,n}(t) &=& {\bf \widetilde{X}}_{n}(t) + {\bf \widetilde{Y}}_{n}(t) + {\bf \widetilde{Z}}_{n}(t) , \\
\label{eq:tadiny}
{\bf R}_{2,n}(t) &=& {\bf \widetilde{X}}_{n}(t) + {\bf \widetilde{Y}}_{n}(t) - {\bf \widetilde{Z}}_{n}(t) , \\
\label{eq:tadinz}
{\bf R}_{3,n}(t) &=& {\bf \widetilde{X}}_{n}(t) - 2{\bf \widetilde{Y}}_{n}(t) .
\end{eqnarray}
\end{subequations}
The dynamics for these modes are very similar as for the modes of the
star polymer. The non-vanishing correlation functions for the modes of
the tadpole are then given by
\begin{subequations}
    \label{eq:tadmacfexact}
    \begin{align}
        \label{eq:tadmacfexacta}
        &\langle [ {\bf X}_{0}(t) - {\bf X}_{0}(0) ]^2 \rangle =
        \frac{2 k_B T}{\zeta (N+1)} t \\
        \label{eq:tadmacfexactb}
        &\langle {\bf X}_{p}(t) \cdot {\bf X}_{q}(0) \rangle =
        \frac{k_B T}{\zeta(N+1)}\frac{1}{2 \alpha_{{\bf X}_{p}}} \exp \left[-\alpha_{{\bf X}_{p}} t\right] \delta_{pq} \\
        \label{eq:tadmacfexactc}
        &\langle {\bf Y}_{p}(t) \cdot {\bf Y}_{q}(0) \rangle =
        \frac{9 k_B T}{4 \zeta (2N+1)}
        \frac{1}{2 \alpha_{{\bf Y}_{p}}} \exp \left[-\alpha_{{\bf Y}_{p}} t\right] \delta_{pq} \\
        \label{eq:tadmacfexactd}
        &\langle {\bf Z}_{p}(t) \cdot {\bf Z}_{q}(0) \rangle =
        \frac{3 k_B T}{\zeta(2N+1)}\frac{1}{2 \alpha_{{\bf Z}_{p}}} \exp \left[-\alpha_{{\bf Z}_{p}} t\right] \delta_{pq} ,
    \end{align}
\end{subequations}
for $p,q=1,\dots N$ and
\begin{align}
\alpha_{{\bf X}_{p}} = 4\frac{k}{\zeta} \sin^2\left[\frac{\pi p}{2N+2}\right] ,
\alpha_{{\bf Y}_{p}} = 4\frac{k}{\zeta} \sin^2\left[\frac{\pi (p-1/2)}{2N+1}\right] ,
\alpha_{{\bf Z}_{p}} = 4\frac{k}{\zeta} \sin^2\left[\frac{\pi p}{2N+1}\right] .
\end{align}

\subsection{Radius of gyration}

When the ends of two arms of a three armed symmetric star polymer are
connected the mobility reduces and the radius of gyration will become
smaller. Following the same steps as for the star polymer the radius
of gyration is defined in the same way for the case when the central
bead is three times as heavy as any other bead, and by plugging in the
inverses it can be rewritten in terms of summations over the
correlation functions of the modes at time zero. The radius of
gyration squared for a long tadpole becomes:
\begin{align}
    \label{eq:gyration7}
    R_{g}^{2} &= \frac{k_B T N}{k} \frac{5}{6},
\end{align}
which is $5/7$th of the radius of gyration of the same polymer but
with the ends not connected.

\subsection{Mean square displacement of a monomer}

Since the tadpole can be seen as a three armed star polymer with ends
of two of the arms connected, the mean square displacement of the
monomers in the third arm and that of the central monomer will not
feel the effect of the connection of the two remaining arms. And so
for the central monomer and the monomers in the tail of the tadpole
the expressions will be exactly the same as in Eqs.
(\ref{eq:starmsd2},\ref{eq:starmsd6}) respectively. A monomer in the
closed circle of the tadpole should behave in the same way as some
monomer roughly in the middle of the tail:
\begin{align}
    \label{eq:tadmsd1}
    \Delta {\bf R}_{0}(t)  \equiv {\bf R}_{0}(t) - {\bf R}_{0}(0)  \quad,\quad
    \Delta {\bf R}_{i,n}(t) \equiv {\bf R}_{i,n}(t) - {\bf R}_{i,n}(0) .
\end{align}
\begin{figure*}
    \centering
    \epsfig{file=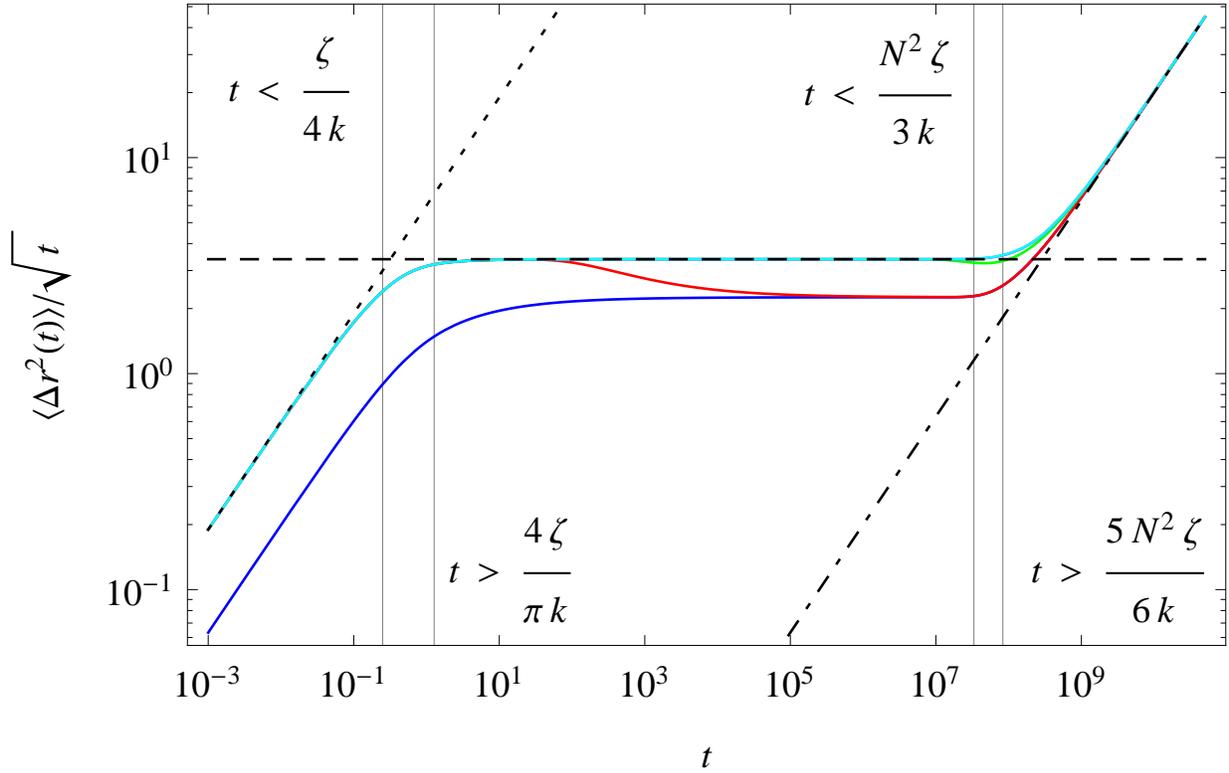,width=1.0\linewidth,clip=}
    \caption{The scaled mean square displacement $\langle \Delta {\bf
        R}^{2}(t) \rangle/\sqrt{t}$ as a function of time. The
      mean square displacement of several beads in a tadpole made by a
      symmetric star polymer with ends of arm $1$ and $2$ connected as
      defined by Eq. (\ref{eq:tadmsd1}) were exactly evaluated
      with $N=10^4$ and other parameters put to $1$. The solid blue
      line corresponds to the mean square displacement of the central
      bead whereas the red, green, magenta, and cyan solid lines
      correspond to $i=1$ and $n=10^{-3} N,N/2,9N/10$, and $N$ in
      Eq. (\ref{eq:tadmsd1}) respectively. Unlike for the unconnected
      arms in the star polymer here there is no real difference
      between a monomer at the middle or end of the connected arm. A
      bead positioned somewhere along the arm will at first behave as
      if it were in the middle of the arm. If it is close enough to
      the heavy central bead the movement will be restricted and the
      mean square displacement will mimic that of the central
      bead. The short time scale $t < \zeta/(4 k)$ and the very long
      time scale $t \gg 5 N^2 \zeta/(6 k)$ for which $\langle \Delta
      {\bf R}^{2}(t) \rangle \sim t$ corresponding to the dotted and
      dot-dashed lines are separated by a time during which $\langle
      \Delta {\bf R}^{2}(t) \rangle \sim \sqrt{t}$ corresponding to
      the dashed line as in agreement with Eqs.
      (\ref{eq:tadmsd}).}
    \label{fig:tadmsd}
\end{figure*}
The mean square displacement for monomers in a tadpole can be
described by
\begin{subequations}
\label{eq:tadmsd}
    \begin{equation}
    \label{eq:tadmsd2}
    \langle \Delta {\bf R}_{0}^{2}(t) \rangle =
    \left\{
    	\begin{array}{ll}
            \displaystyle{\frac{2 k_B T}{\zeta} t, \quad\quad\quad\quad\mbox{for}\, t<\frac{\zeta}{4 k}}\\
            \displaystyle{ 4 k_B T \sqrt{\frac{t}{\pi k \zeta}},
              \quad\,\,\,\,\mbox{for}\, \frac{4 \zeta}{\pi k} < t < \frac{N^2 \zeta}{3 k}}\\
            \displaystyle{\frac{2 k_B T}{\zeta N} t,
              \quad\quad\quad\quad\mbox{for}\, t > \frac{N^2 \zeta}{3 k}}
    	\end{array}
    \right.,
    \end{equation}
\begin{equation}\label{eq:tadmsd3}
\langle \Delta {\bf R}_{3,N}^{2}(t) \rangle =
\left\{
	\begin{array}{ll}
        \displaystyle{\frac{6 k_B T}{\zeta} t, \quad\quad\quad\quad\mbox{for}\, t<\frac{\zeta}{4 k}}\\
        \displaystyle{ 12 k_B T\sqrt{\frac{t}{\pi k \zeta}}, \quad\,\,\,\,\mbox{for}\, \frac{4 \zeta}{\pi k} < t < \frac{N^2 \zeta}{3 k}}\\
        \displaystyle{\frac{2 k_B T}{\zeta N} t, \quad\quad\quad\quad\mbox{for}\, t > \frac{7 N^2 \zeta}{3 k}}
	\end{array}
\right.,
\end{equation}
\begin{equation}\label{eq:tadmsd4}
\langle \Delta {\bf R}_{1,N}^{2}(t) \rangle = \langle \Delta {\bf R}_{3,N/2}^{2}(t) \rangle =
\left\{
	\begin{array}{ll}
        \displaystyle{\frac{6 k_B T}{\zeta} t, \quad\quad\quad\quad\mbox{for}\, t<\frac{\zeta}{4 k}}\\
        \displaystyle{ 6 k_B T\sqrt{\frac{t}{\pi k \zeta}}, \quad\,\,\,\,\mbox{for}\,\frac{4 \zeta}{\pi k} < t < \frac{N^2 \zeta}{3 k}}\\
        \displaystyle{\frac{2 k_B T}{\zeta N} t,
          \quad\quad\quad\quad\mbox{for}\, t > \frac{5 N^2 \zeta}{6 k}}
	\end{array}
\right..
\end{equation}
\end{subequations}
Note that the domains of validity for the different behavior have
changed in comparison to the star polymer. The exact sums for the
monomers in a tadpole have a slightly different maximum as can be
calculated using Eq. (\ref{eq:maxsum}).

\section{Discussion\label{disc}}

For bead-spring models of polymers with the topology
of a symmetric star with $f$ arms,
where the hydrodynamic radius of the central bead is $f$ times
as heavy as any other bead, we derived the exact expressions for the
dynamical eigenmodes. We demonstrated the usefulness of this exercise 
by exact calculations of various quantities that yield prefactors as well 
as give insights into the scaling behavior at different time regimes 
for these quantities. The radius of gyration was shown to be given by 
$R_{g}^{2}= k_B T N (3f-2)/ (2kf)$; the mean square displacement of the 
central bead and of various other individual beads was shown to scale as 
$\sim t$ at very short as well as at very long times, with an intermediate 
regime in which it scales as $\sim \sqrt{t}$; and the correlation 
function of an orientational vector was shown to stay invariant 
at very short times, decay exponentially at very long times, and 
decay as $\sim 1/\sqrt{t}$ in the intermediate regime. Similar results were
also derived for tadpoles, which are $f=3$ star polymers in which two
arms are connected.

Our exact results are limited to the very specific case of $f$-arm star polymers and tadpoles
in which the hydrodynamic radius of the central bead is exactly $f$ times bigger 
than the other beads. The Rouse modes can still be constructed for other cases, 
since the Rouse equations are linear in the bead positions; indeed, this has 
already been achieved numerically\cite{starthesis}. We do not rule out analytical 
forms of the Rouse modes for these cases, although we can expect that these 
analytical forms may not look aesthetically pleasant. Even if the hydrodynamic 
radius of the central bead is not exactly $f$ times bigger than the other beads, 
we do not forsee differences in the behavior of the mean-square displacements in 
the scaling limit, namely the three regimes mentioned above, albeit with different
prefactors. The same should hold for modest modifications of the interatomic bonds: 
no qualitatively different behavior is expected, but prefactors will be affected. 
As for the dynamical eigenmodes of star and tadpole polymers, qualitative 
differences will however arise due to significant hydrodynamic interactions 
between the beads, excluded-volume effects, or if the solution is no longer dilute.

Apart from these direct results, the usefulness of our exercise lies
perhaps even more in providing a proper language in which to
characterize the dynamics of star and tadpole polymers; very similar
to the ubiquitous use of Rouse mode amplitudes for linear polymers.

In future work, we will use the mathematical expressions for the Rouse modes
derived here, to characterize the dynamics of star polymers in simulations
of self-avoiding stars, in dilute circumstances, as well as in various dense
environments (such as a melt of linear polymers, or a homodisperse melt of
star polymers).

\appendix

\section{Determination of the Rouse modes of a star polymer}
\label{sec:math}
Several steps are needed to derive the dynamics of the modes (\ref{eq:starmacf})
from the equations of motion for a symmetric star polymer. It is
useful to note that the modes are complete, which allows us to express
the positions of the beads from the mode amplitudes, given by
\begin{subequations}
    \label{eq:starinv}
    \begin{align}
        \label{eq:starinva}
        {\bf R}_{0} =& {\bf X}_{0} + 2 \sum_{p=1}^{N} \cos\left[\frac{\pi p/2}{N+1}\right] {\bf X}_{p} \\
        {\bf R}_{i,n} =& {\bf X}_{0} + 2 \sum_{p=1}^{N} \cos\left[\frac{\pi (n+1/2) p}{N+1}\right] {\bf X}_{p}+\frac{4}{f} \sum_{j,p=1}^{f,N}
        \cos\left[\frac{\pi(N-n+1/2)(p-1/2)}{N+1/2}\right] {\bf
          Y}_{p}^{(i,j)}.
\label{eq:starinvb}
    \end{align}
\end{subequations}
The correctness of these equations can be checked using the
orthogonality relations (\ref{eq:ortho}).

The dynamical equations of motion for the star polymer combined with
the potential energy then result in the following equation of motion
for the beads:
\begin{subequations}
    \label{eq:stareom}
    \begin{align}
        \label{eq:stareoma}
        \frac{d {\bf R}_{0\,\,\,\,}}{d t} &= -\frac{k}{\zeta} \left( {\bf R}_{0} - \frac{1}{f }\sum_{i=1}^{f} {\bf R}_{i,1} \right) +
        {\bf g}_{0} \\
        \label{eq:stareomb}
        \frac{d {\bf R}_{i,1}}{d t} &= -\frac{k}{\zeta} \left( 2 {\bf R}_{i,1} - {\bf R}_{0} - {\bf R}_{i,2} \right)
        + {\bf g}_{i,1} \\
        \label{eq:stareomc}
        \frac{d {\bf R}_{i,n}}{d t} &= -\frac{k}{\zeta} \left( 2 {\bf R}_{i,n} - {\bf R}_{i,n-1} - {\bf R}_{i,n+1} \right) + {\bf g}_{i,n} .
    \end{align}
\end{subequations}
Note that $i=1\dots f$ and Eq. (\ref{eq:stareomc}) is valid for
all $n=2\dots N$ where Eq. (\ref{eq:starinvb}) is needed to show
that ${\bf R}_{i,N+1}={\bf  R}_{i,N}$. For convenience we define
\begin{equation}
\label{eq:temp1}
    {\bf R}_{n} = \frac{1}{f} \sum_{i=1}^{f} {\bf R}_{i,n} = {\bf X}_{0} + 2 \sum_{p=1}^{N} \cos\left[\frac{\pi (n+1/2) p}{N+1}\right] {\bf X}_{p} ,
\end{equation}
for $n=0 \dots N$ which coincides with the inverse of ${\bf R}_{0}$ as
given in Eq. (\ref{eq:starinva}). Taking the time derivative on both
sides of Eq. (\ref{eq:starxmode}), plugging in the equations of
motion for the beads from Eq. (\ref{eq:stareom}), and using the
definition above for which ${\bf R}_{-1}={\bf R}_{0}$ and ${\bf
  R}_{N+1}={\bf R}_{N}$ results in
\begin{equation}
\label{eq:temp2}
    \frac{d {\bf X}_{p}}{d t}=-\frac{1}{N+1}\frac{k}{\zeta} \sum_{n=0}^{N} \cos\left[\frac{\pi (n+1/2) p}{N+1}\right] \left(2 {\bf R}_{n} - {\bf R}_{n-1} - {\bf R}_{n+1} \right) + {\bf G}_{p} \,\,,
\end{equation}
where ${\bf G}_{p}$ is the transform of the thermal forces which will
be calculated later. By using (\ref{eq:temp1}), the trigonometric identities, namely the
angle sum and difference identities and the power-reduction formula,
and finally the orthogonality relation the set of differential
equations becomes
\begin{align}
    \label{eq:starxeom}
    \frac{d {\bf X}_{p}}{d t} = \left\{
    \begin{array}{ll}
        - \alpha_{{\bf X}_{p}} {\bf X}_{p} + {\bf G}_{p},\quad p=1 \dots N \\
        {\bf G}_{0} \qquad\qquad\quad, \quad p=0
    \end{array} \right.
    ,\quad\alpha_{{\bf X}_{p}} = 4\frac{k}{\zeta} \sin^2\left[\frac{\pi p}{2N+2}\right] .
\end{align}
The set of differential equations for the ${\bf Y}^{(i,j)}_{p}$ modes
can be written down in a similar manner. We define ${\bf
  R}_{n}^{(i,j)} \equiv {\bf R}_{i,n}-{\bf R}_{j,n}$ and follow
similar steps as for the ${\bf X}_{p}$ modes so that
\begin{align}
    \label{eq:staryeom}
    \frac{d {\bf Y}_{p}^{(i,j)}}{d t} = -\alpha_{{\bf Y}_{p}} {\bf Y}_{p} + {\bf G}_{p}^{(i,j)}, \quad p=1 \dots N,
    \quad\alpha_{{\bf Y}_{p}} \equiv 4\frac{k}{\zeta} \sin^2\left[\frac{\pi (p-1/2)}{2N+1}\right]
\end{align}
for some transform of the thermal forces ${\bf G}_{p}^{(i,j)}$.

With the above transformations, the set of differential equations for
the beads have been transformed to a set of disconnected linear
differential equations.

To find the relation for the modes as in Eq. (\ref{eq:starmacf})
we must first determine the transform of the thermal forces ${\bf
  G}_{p}$ and ${\bf G}_{p}^{(i,j)}$. The transforms are equal to that
of the modes in Eq. (\ref{eq:starmode}) but by replacing the
position of the beads by their thermal force. Use that the thermal
forces are uncorrelated in time and between beads as in Eq.
(\ref{eq:thermal}) and recall that the central bead has a friction
coefficient $f$ times as large as that of the other beads. The only
nonvanishing functions with $p=1 \dots N$ are
\begin{subequations}
    \label{eq:tcfg}
    \begin{align}
        \label{eq:tcfga}
        \left\langle {\bf G}_{0}(t) \cdot {\bf G}_{0}(t\sp{\prime}) \right\rangle &=
        \frac{6 k_B T}{\zeta f(N+1)} \delta(t-t\sp{\prime}) \\
        \label{eq:tcfgb}
        \left\langle {\bf G}_{p}(t) \cdot {\bf G}_{q}(t\sp{\prime}) \right\rangle &=
        \frac{3 k_B T}{\zeta f(N+1)} \delta_{p q} \delta(t-t\sp{\prime}) \\
        \label{eq:tcfgc}
        \left\langle {\bf G}_{p}^{(i,j)}(t) \cdot {\bf G}_{q}^{(k,l)}(t\sp{\prime}) \right\rangle &=
        \frac{3 k_B T}{\zeta(2N+1)} \delta_{p q} \delta(t-t\sp{\prime})
        \frac{\delta_{(i,j),(k,l)}}{2}.
    \end{align}
\end{subequations}
With these, the set of differential equations can be solved exactly
resulting in the following relations between modes:
\begin{subequations}
    \label{eq:starmacfexact}
    \begin{align}
        \label{eq:starmacfexacta}
        &\langle [ {\bf X}_{0}(t) - {\bf X}_{0}(0) ]^2 \rangle =
        \frac{6 k_B T}{\zeta f(N+1)} t \\
        \label{eq:starmacfexactb}
        &\langle {\bf X}_{p}(t) \cdot {\bf X}_{q}(0) \rangle =
        \frac{3 k_B T}{\zeta f(N+1)}\frac{1}{2 \alpha_{{\bf X}_{p}}} \exp \left[-\alpha_{{\bf X}_{p}} t\right] \delta_{pq} \\
        \label{eq:starmacfexactc}
        &\langle {\bf Y}_{p}^{(i,j)}(t) \cdot {\bf Y}_{q}^{(k,l)}(0) \rangle =
        \frac{3 k_B T}{\zeta (2N+1)}
        \frac{\delta_{(i,j)(k,l)}}{2} \frac{1}{2 \alpha_{{\bf Y}_{p}}} \exp \left[-\alpha_{{\bf Y}_{p}} t\right] \delta_{pq} ,
    \end{align}
\end{subequations}
where all other correlations between modes are strictly zero. By
taking the long-polymer limit the sines can be expanded up to second
order and the results are in Eq. (\ref{eq:starmacf}).

\section{Useful mathematical relations}

The following relations are useful for calculating the Rouse modes:
\begin{subequations}
\label{eq:ortho}
\begin{align}
    \label{eq:ortho1}
  &\frac{2}{N+1}\sum_{n=0}^{N} \cos\left[\frac{\pi(n+1/2)p}{N+1}\right] \cos\left[\frac{\pi(n+1/2)q}{N+1}\right] = \delta_{pq}
\\
    \label{eq:ortho2}
    &\frac{4}{2N+1} \sum_{n=1}^{N} C_p C_q = \delta_{pq}
    \quad,\quad
    C_p = \cos\left[\frac{\pi(N-n+1/2)(p-1/2)}{N+1/2}\right],
\\
    \label{eq:ortho3}
    &\frac{4}{2N+1} \sum_{n=1}^{N} C_p C_q = \delta_{pq}
    \quad,\quad
    C_p = \sin\left[\frac{\pi(N-n+1/2)p}{N+1/2}\right],
\end{align}
\end{subequations}
for $p,q=1\dots N$ and where the left hand side of Eq.
(\ref{eq:ortho1}) equals $2 \delta_{p 0}$ for $p=0\dots N$.

\begin{subequations}
\label{eq:minsum}
\begin{align}
\label{eq:minsum1}
&\sum_{p=1}^{N} \cos^2 \left[ \frac{\pi(n+1/2)p}{N+1} \right] = \frac{N}{2} , \quad n=0 \dots N\\
\label{eq:minsum2}
&\sum_{p=1}^{N} \cos^2 \left[ \frac{\pi(N-n+1/2)(p-1/2)}{N+1/2} \right] = \frac{2N+1}{4} , \quad n=1 \dots N \\
\label{eq:minsum3}
&\sum_{p=1}^{N} \sin^2 \left[ \frac{\pi(N-n+1/2)p}{N+1/2} \right] = \frac{2N+1}{4} , \quad n=1 \dots N
\end{align}
\end{subequations}

\begin{subequations}
\label{eq:maxsum}
\begin{align}
\label{eq:maxsum1}
&\sum_{p=1}^{N} \cos^2 \left[ \frac{\pi(n+1/2)p}{N+1} \right]\sin^{-2} \left[ \frac{\pi p}{2N+2} \right] = \frac{N}{3}(2N+1)-2N n+2n^2
\\
\label{eq:maxsum2}
&\sum_{p=1}^{N} \cos^2 \left[ \frac{\pi(N-n+1/2)(p-1/2)}{N+1/2} \right] \sin^{-2} \left[ \frac{\pi (p-1/2)}{2N+1} \right] = (2N+1)n
\\
\label{eq:maxsum3}
&\sum_{p=1}^{N} \sin^2 \left[ \frac{\pi(N-n+1/2)p}{N+1/2} \right] \sin^{-2} \left[ \frac{\pi p}{2N+1} \right] = (2N+1-2n)n
\\
\label{eq:maxsum4}
&\sum_{p=1}^{N} \sin^2 \left[ \frac{\pi(n+1)p/2}{N+1} \right] \sin^2 \left[ \frac{\pi n p/2}{N+1} \right] \sin^{-2} \left[ \frac{\pi p/2}{N+1} \right] = \frac{n}{2}(N+1)
\end{align}
\end{subequations}

\begin{subequations}
    \label{eq:sums}
    \begin{align}
        \label{eq:sumsa}
        &\sum_{p=1}^{\infty} \frac{1}{p^{2}} = \frac{\pi^2}{6} \\\
        \label{eq:sumsb}
        &\sum_{p=1}^{\infty} \frac{1}{(p-1/2)^{2}} = \frac{\pi^2}{2} \qquad
    \end{align}
\end{subequations}

\end{document}